\documentstyle[11pt,saltwshop,twoside,epsf]{article}
\def\edcomment#1{\iffalse\marginpar{\raggedright\sl#1\/}\else\relax\fi}
\reversemarginpar

\begin{document}

\title{Chemical and mass evolution of galaxies}
 \author{Bodo L. Ziegler}
\affil{Universit\"atssternwarte G\"ottingen, Geismarlandstr. 11, 37083
G\"ottingen, Germany}
%\author{Ima Co-Author}
%\affil{The Name of My Institution, The Full Address of My Institution}

\begin{abstract}
An introduction is given to projects investigating galaxy evolution 
quantitatively by spectroscopic observations of very distant galaxies that 
have weak apparent brightnesses and small sizes as it is feasible with 
10m-class telescopes like SALT.
Such methods encompass scaling relations like the Tully--Fisher and
Fundamental Plane relations that can be utilized to determine the luminosity
evolution and mass assembly of galaxies.
The stellar populations can be analyzed with respect to age, metallicity, and
chemical enrichment by measureing absorption line strengths.
Possible effects on galaxy evolution of the environment in rich clusters of 
galaxies compared to the field are also addressed.
For each method, recent applications are presented like the evolution of the
TFR determined with 77 field spirals up to $z=1$, a study of the internal
kinematics of distant cluster spirals and a comparison of the stellar 
populations of ellipticals in the field and in rich clusters at $z=0.4$.
\end{abstract}

\section{\label{intro}Introduction}

In the local Universe we see a variety of galaxy types that can be 
distinguished both by their morphology and their internal kinematics.
The two main families are the spiral galaxies in which stars and gas rotate
around the center in a disk and the elliptical galaxies that are 
gravitationally stabilized by the random motion of the stars within a triaxial
body.
Some kind of intermediate class are the lenticular or S0 galaxies with central
bulges surrounded by disks.
In addition, we observe irregular galaxies with many different structures.

One of the main tasks of extragalactic research is to understand why the
galaxies appear in such different types. 
Is the structure of a galaxy already fixed at the time of its formation?
Or does it change during the time of its evolution and is it subject to 
external influence?
The vastly accepted paradigm of galaxy formation is nowadays the theory of
the hierarchical growth of all structure in the Universe.
Large galaxies are built up via the merging of smaller systems.
In such events, the disk structure of the progenitors is transformed into a 
spheroidal shape.
An important ingredient of the theory is the dominance of cold dark matter
and the main prediction is that the total mass of galaxies is growing
during the evolution.

A first step to quantify galaxy evolution is to study the luminosity of 
galaxies up to high redshifts with deep photometric observations.
Well-known deep imaging are the Hubble Deep Fields 
(Williams et al. 1996; Casertano et al. 2000)
but their major drawback is their very small field size.
Therefore, the F{\sc ors} consortium\footnote{Institutes in G\"ottingen,
Heidelberg, and Munich under the PI Prof. Appenzeller.}
made the effort to observe a comparable deep but larger field with the 
F{\sc ors} instrument at the VLT. 
Consuming some 72 hours pure integration time, deep images were taken in the
$UBgRIz$ filters, supplemented by $J$ and $K$ near-infrared exposures with
the NTT (Heidt et al. 2003).
With the aid of photometric redshifts, number counts and luminosity functions
can be constructed up to redshifts $z\approx5$ in this F{\sc ors} Deep Field
(FDF; Gabasch et al. 2004).

But a more robust test of the Cold Dark Matter theory would be the measurement
of the mass and size of distant galaxies, because the brightness of galaxies
is subject to star formation processes. 
A rather quiescent luminosity evolution of the stellar population of a galaxy
can get quickly wiped out by any contemporary star burst even if it 
involves only a small fraction in mass.
From pure photometry it is very difficult to judge whether any galaxy within
a sample is undergoing such an event.

With 10m-class telescopes it is nowadays possible to take spectra with
sufficient spectral and spatial resolution of faint galaxies down to
$\sim 23 R$ mag reaching the necessary $S/N$ within reasonable exposure times.
Analyzing spectral lines, the characteristic velocity of the stars and gas 
within a galaxy can be determined which is a measure for its total mass.
For elliptical galaxies, the width of absorption lines reflects the internal
velocity dispersion, while for spiral galaxies, the Doppler shifts of emission
lines trace the rotational speed.
The characteristic size of distant galaxies can be determined in spatially
highly resolved images as they can be obtained either with the HST or
groundbased telescopes equipped with adaptive optics.
Important tools for the quantitative investigation of galaxy evolution are
scaling relations because entire samples of galaxies can be compared with each
other rather than just individual galaxies.
Such scaling relations are, for example, the Fundamental Plane of ellipticals
(e.g. Bender, Burstein, \& Faber 1992) 
or the Tully--Fisher relation (TFR) of spirals 
(e.g. Pierce \& Tully 1992 [PT92]).

\section{Evolution of the Tully--Fisher relation}

As one example, the TFR of galaxies up to $z\approx1$ corresponding to a
lookback time of $\sim8$\,Gyr or half the age of the Universe\footnote{The 
``concordance'' cosmology is used
with $\Omega_{\rm matter}=0.3$, $\Omega_{\lambda}=0.7$, and 
$H_0 = 70$\,km\,s$^{-1}$\,Mpc$^{-1}$.} 
is shown in Figure\,1.
Only emission--line galaxies that exhibit a rotation curve (RC) which turns 
into a flat regime with constant velocity are suitable for display in 
such a TF diagram. 
The maximum rotational speed ($v_{\rm max}$) in such a case is then a measure 
for the total mass of the galaxy including its Dark Matter. 
Distant galaxies impose difficulties on the analysis of RCs not only due to 
their faintness but because their small sizes limit strongly the spatial 
extent of any emission line.
From a sample of 113 spirals in the FDF, we were able to determine 
$v_{\rm max}$ for 77 galaxies with $0.1\le z\le1$.

\begin{figure}
\plottwo{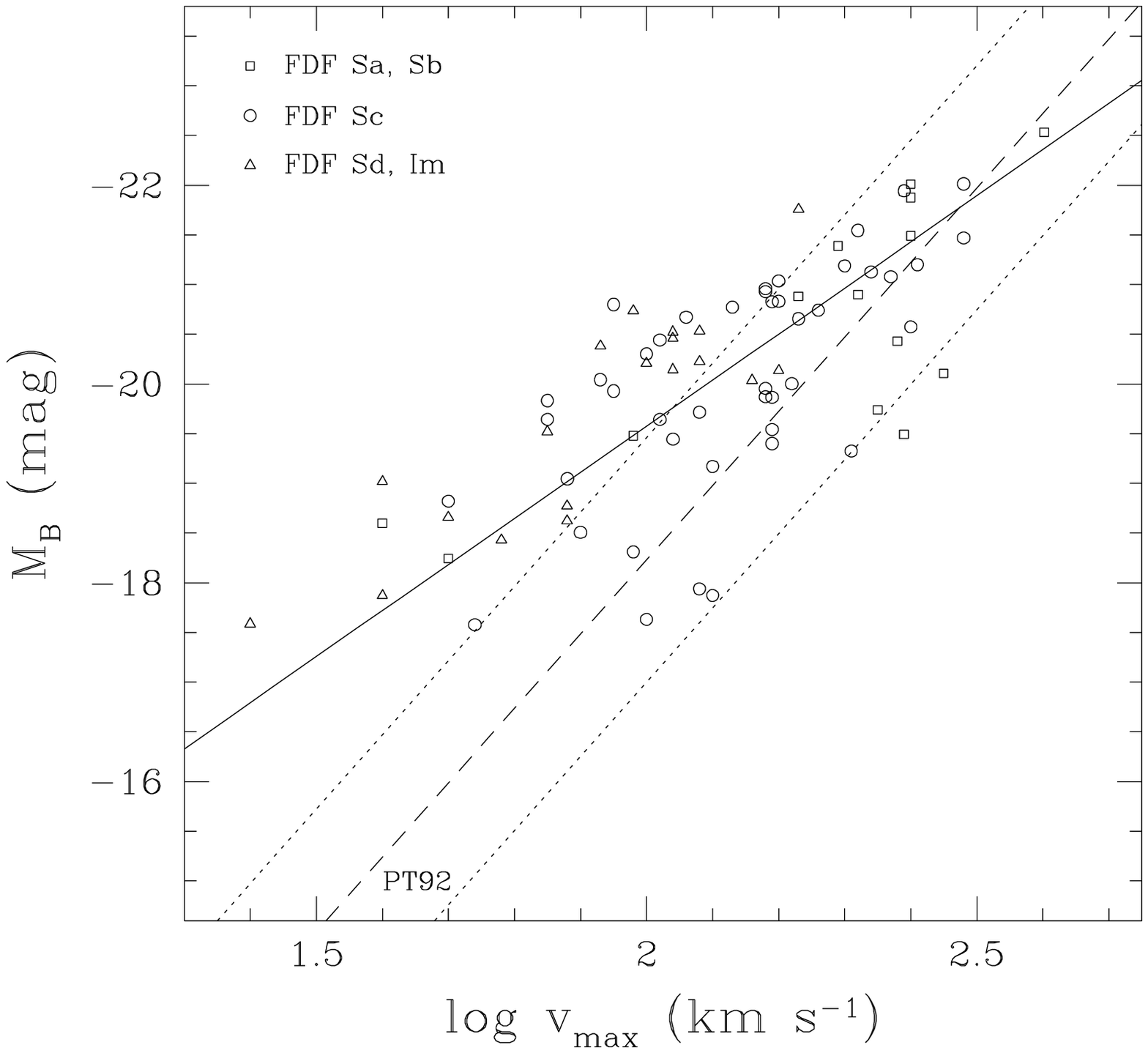}{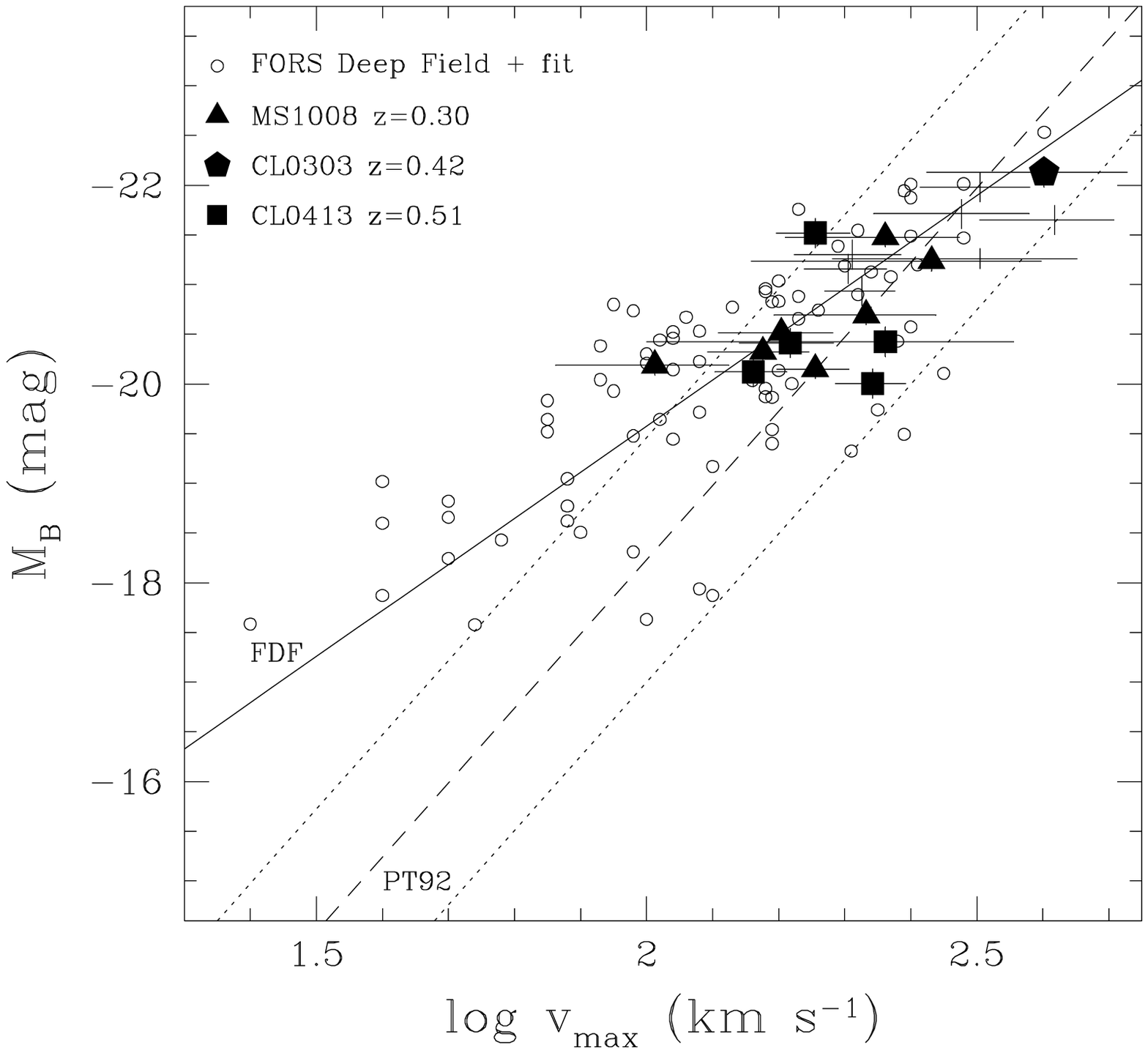}
\caption{
Panel a) The TFR of 77 distant FDF field spirals compared to the local 
relation ($\pm3\sigma$) from [PT92]. 
The distant sample is subdivided into types Sa \& Sb, Sc, and Sd \& Im.
Panel b) 13 galaxy members of distant clusters overplotted on the FDF field
sample.
}
\end{figure}

Comparing the restframe $B$ magnitudes for fixed $v_{\rm max}$ between the
distant FDF sample and a local sample by PT92, we find a substantial increase
in the luminosity for small galaxies but very similar brightnesses for large
ones (Ziegler et al. 2002; B\"ohm et al. 2004).
Such a mass-dependent evolution is not easily explained by CDM theory
since it involves the modeling of stellar populations.
First simulations indicate that the efficiency, strength and duration of star 
formation in a galaxy may depend on its total mass.
For a better understanding, spectroscopic observations of many more distant
galaxies are needed, too, as it will be possible with SALT.
This becomes even more obvious when we try to subdivide the emission--line 
galaxies into different types as in Figure\,1a.
Looking into such subsamples will allow to investigate the impact of
varying global parameters like star formation rates or mass-to-light ratios.

\section{Environmental dependence of galaxy evolution}

Looking at the distribution of different galaxy types, it becomes obvious that
the average mean distance between galaxies must play a role in shaping a 
galaxy's structure. 
This morphology--density relation manifests itself most clearly in the 
prevalence of early--type galaxies in the center of rich clusters of galaxies 
(Dressler 1980). 
At the same time, interactions between cluster galaxies and the hot X-ray 
emitting intracluster gas or the cluster potential indicate phenomena specific
to clusters (e.g. van Gorkom 2003).
We also observe evolution in the galaxy population of clusters, i.e. the
mixture of the various types.
Distant clusters at $z\approx0.5$ have, for example, a greatly reduced 
fraction of S0 galaxies that dominate the bright population locally while the 
number of ellipticals stays nearly constant (Dressler et al. 1997).
These findings have provoked the question whether field spirals falling into
a cluster could get transformed into S0 galaxies during the time due to
clusterspecific processes.
Another environmental dependence is given for the global star formation as it 
is observed to be largely suppressed in cluster members 
(e.g. G\'omez et al. 2003).
Investigating galaxies in densities intermediate between the field and cluster
environments reveals that at least two different interaction processes on
different timescales must be responsible for transforming galaxy structure
and suppressing star formation, respectively (e.g. Balogh et al. 2002).

Interactions can also affect a galaxy's internal kinematics creating asymmetric
RCs, for example, or totally corrupted velocity fields 
(e.g. Rubin, Waterman, \& Kenney 1999).
In a Universe with hierarchically growing structure, it is predicted that the
galaxy merger rate and frequency of infalling galaxies increases with redshift.
Observations should therefore investigate the internal kinematics of galaxies
in clusters up to $z\approx1$ as it is now possible with the big telescopes 
like SALT.
Indeed, spirals in clusters with $0.3<z<0.6$, which we studied with F{\sc ors},
display peculiar RCs (Figure\,2; Ziegler et al. 2003).
On the other hand, several cluster members have undistorted RCs with an outer
flat part.
These galaxies can be compared to field galaxies in a TF diagram allowing to
investigate whether they might be under- or over-luminous for their measured
$v_{\rm max}$.
Truncation of star formation in a cluster member leads to a dimming, whereas
interaction-induced star bursts results in a brightening of the galaxy.
Our cluster targets are distributed very similar to the distant FDF spirals,
so that their stellar populations are momentarily not suspect to any changes
(Figure\,1b).
But it may well be that those galaxies with distorted RCs in our sample have
star formation rates which got affected by the same interaction process that
produced the kinematic peculiarity.
Much larger samples of distant cluster galaxies are needed to robustly
investigate these issues.

\begin{figure}
\plotone{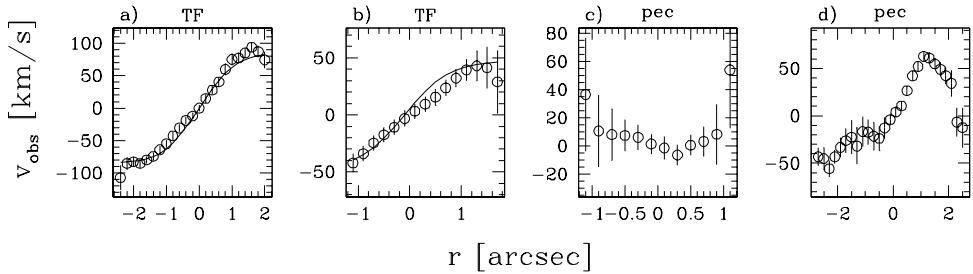}
\caption{Examples of rotation curves of galaxies in distant clusters. Panels a)
and b) represent RCs turning into a flat part where $v_{\rm max}$ can be 
measured by modeling the synthetic velocity field (solid line). Panels c) and
d) display peculiar velocity curves.}
\end{figure}

\section{Evolution of the stellar populations}

Another important tool next to scaling relations for the quantitative 
measurement of galaxy evolution are diagnostic diagrams where absorption line
strengths are plotted versus each other and compared to a model grid.
A widely used application is the Lick/IDS system, in which Balmer lines (in
particular H$\beta$) are taken as age sensitive indicators whereas the
metallicity ([Z/H]) is measured by combining Mg and Fe lines 
(e.g. Faber et al. 1985).
More sophisticated stellar population models can nowadays also distinguish
between different enrichment histories for various elements and are no longer
fixed to the solar abundance (e.g. Thomas, Maraston, \& Bender 2003 [TMB03]).
This opens the possibility for another important test of CDM theory.
It is predicted that large elliptical galaxies in high-density peaks 
corresponding to rich clusters are formed in a very short time ($<1$\,Gyr)
with the bulk of their stars created in a vigorous star burst, whereas
ellipticals in the field get assembled all the time by major mergers of spirals
that already had an extended star formation history
(e.g. Kauffmann 1996).
The stellar populations of ellipticals in different environments should 
therefore have different enrichment histories for the Mg or $\alpha$ elements
and the Fe-peak elements which can only be measured by spectroscopy.
This comes about, because these element families are to first order mainly 
produced by supernovae of type II and type Ia, respectively, which have very 
different timescales.
SN\,IIs are exploded massive stars which have a lifetime of only a few Myrs.
SN\,Ia are the endproducts of a binary stellar system with an evolutionary
time of $\sim1$\,Gyr.
If a cluster elliptical ended its star burst within a few Myrs, then it should
be deficient in Fe leading to Mg/Fe ratios much larger than the solar mix or
than in field ellipticals.
Since an accurate measurement of the absorption line strengths needs a minimum
S/N of $\sim30$, spectroscopy of distant faint galaxies is only feasible with
the 10m-class telescopes like SALT.

\begin{figure}
\plottwo{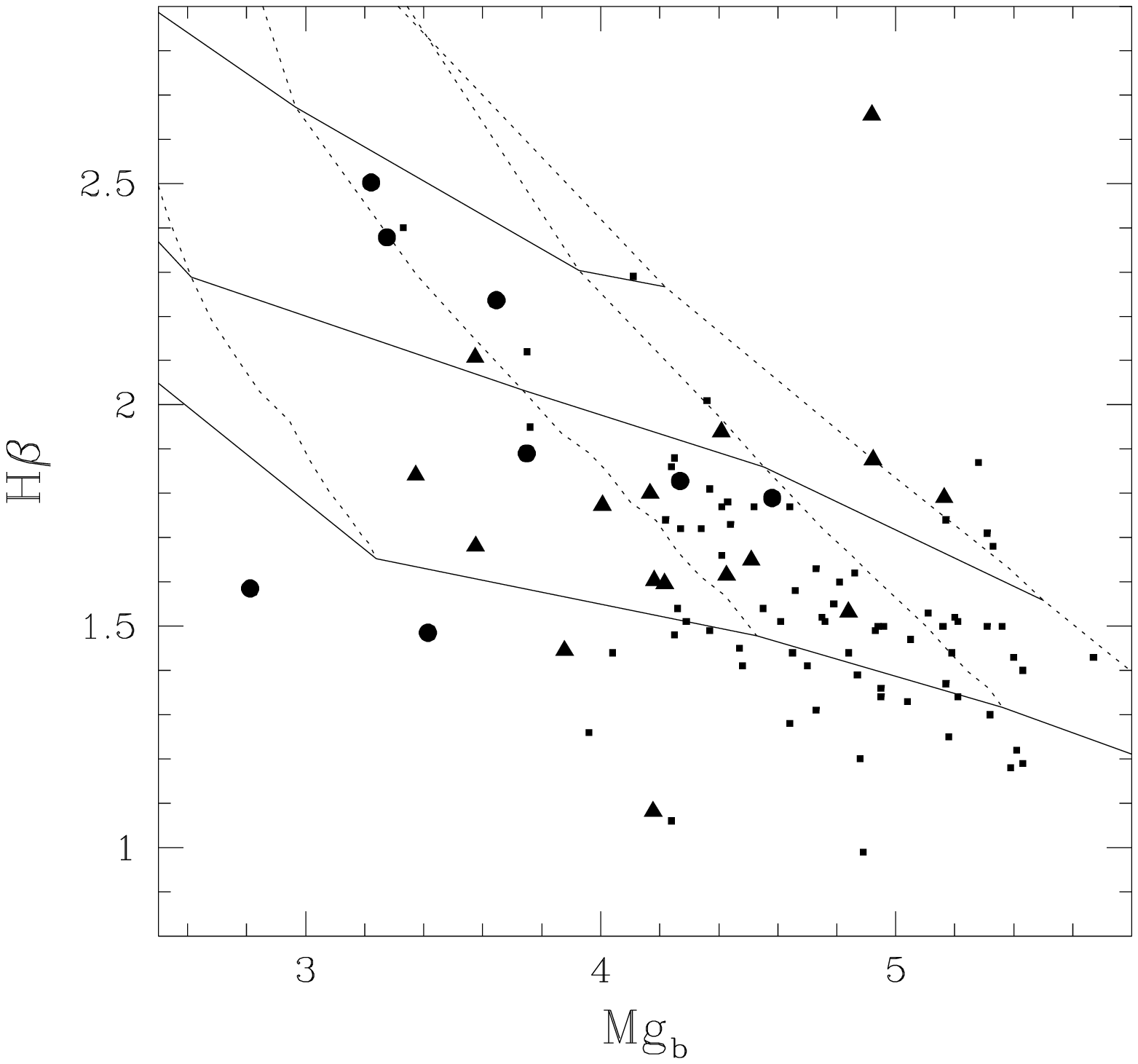}{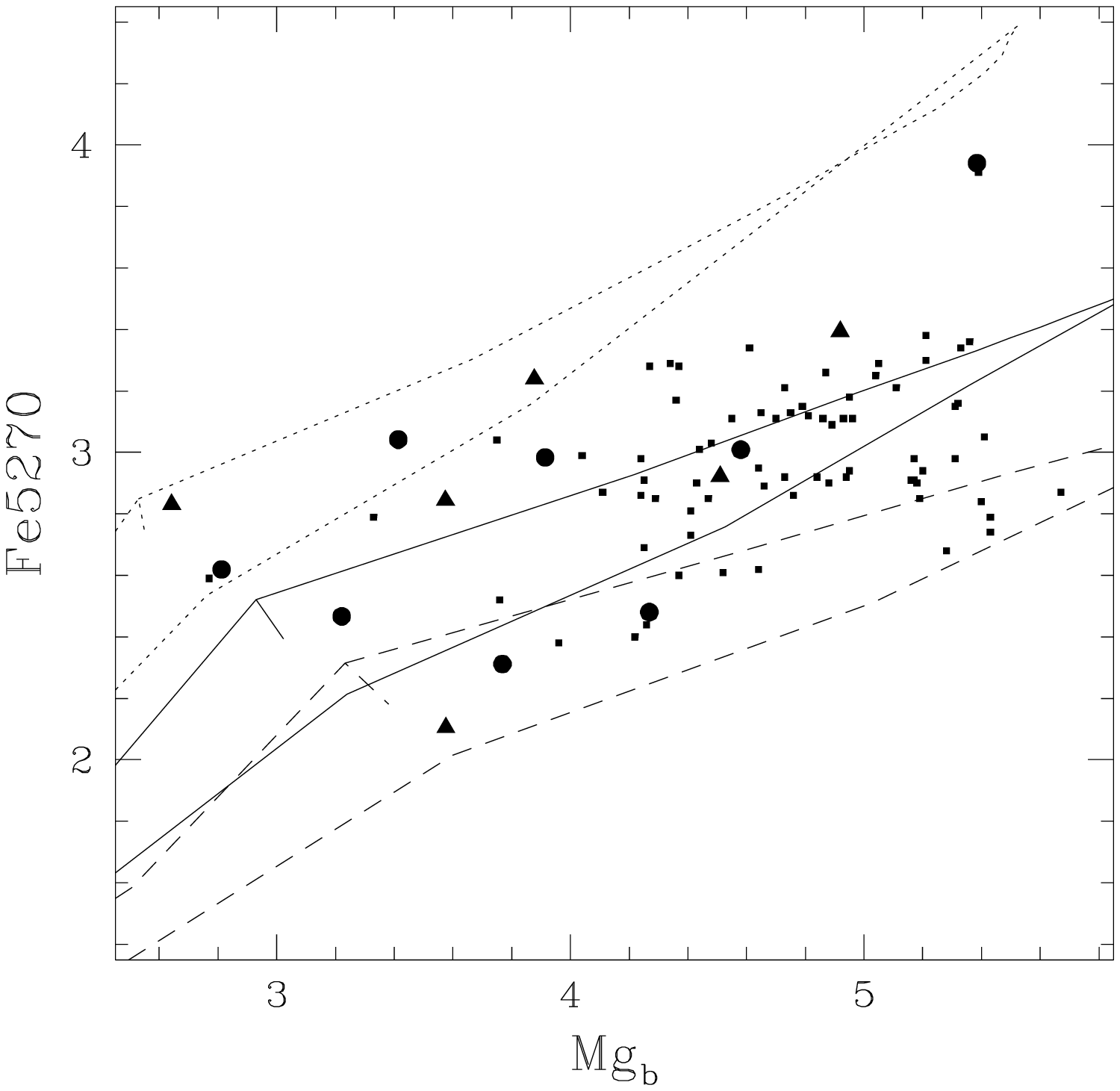}
\caption{
Panel a) Age--metallicity diagram for field (circles) and cluster 
(triangles) ellipticals with $\langle z\rangle\approx0.4$ compared to a sample
of Coma cluster galaxies (small squares; from [K01]). 
Overplotted are [TMB03]-models enhanced in Mg ([Mg/Fe] $= 0.3$) with ages 
1, 5 \& 15\,Gyr (solid lines, top to bottom) and 
[Z/H] $= -0.3,0,0.3,0.7$ (dotted lines, left to right).
Panel b) The same samples with models separated by [Mg/Fe] $= 0,0.3,0.5$
(top to bottom).
}
\end{figure}

As an example, two diagnostic diagrams are shown in Figure\,3 for early--type
galaxies in the Coma cluster (Kuntschner et al. 2001) together with field
ellipticals from the FDF (Ziegler et al. 2004) and cluster ellipticals 
(Ziegler \& Bender 1997) both samples having $\langle z\rangle\approx0.4$.
% A model grid from [TMB03] is overplotted.
Since absorption lines of distant galaxies can be corrupted if they are
redshifted to a region with strong terrestrial emission or absorption, the 
number
of galaxies is different for the two panels, and we do not use the classical
combination of Fe lines but only the Fe5270 index for the assessment of Fe 
abundance.
The bulk of the data in Panel\,a) can be modeled with old ages with the distant
galaxies being younger by $\sim4$\,Gyr on average than Coma compatible with 
the lookback time.
Three of the FDF field objects have luminosity-weighted ages of $\sim2$\,Gyr
and solar metallicity which may indicate that these galaxies are the
endproducts of a recent merger between spirals.
None of the distant field ellipticals reach the peak values in metallicity 
of the cluster members.
The majority of the $z\approx0.4$ galaxies exhibit a super-solar Mg 
overabundance relative to Fe like the local reference (Panel b).

\acknowledgments
Most of the work presented here stems from my Volks\-wagen Foundation Junior
Research Group ``Kinematic Evolution of Galaxies'' that includes Dr. I.
Berentzen, Dr. A. B\"ohm, A. Fritz, B. Gerken, and Dr. K. J\"ager.
I fully acknowledge their permission to show the results of their research
here.
I also acknowledge the very fruitful collaborations within the F{\sc ors} 
consortium and with the University of Durham.
This work has been supported by the Volkswagen Foundation (I/76\,520).
%and the Deutsche Forschungsgemeinschaft (Fr 325/46--1 and SFB 439).

\end{document}